\documentclass[final,5p,times,twocolumn]{my_elsarticle}

\usepackage{amssymb}

\begin{document}

\begin{frontmatter}



\title{Potential of a next generation neutrinoless double beta decay experiment based on ZnMoO$_4$ scintillating bolometers}


\author[BER]{J.W.~Beeman} 
\author[INR]{F.A.~Danevich}
\author[NTS]{V.Ya.~Degoda}
\author[NVS]{E.N.~Galashov}
\author[ORS,INS]{A.~Giuliani\corref{cor1}}
\cortext[cor1]{Corresponding author.}
\ead{andrea.giuliani@csnsm.in2p3.fr}
\author[INR]{V.V.~Kobychev} 
\author[INS,ORS]{M.~Mancuso}
\author[ORS]{S.~Marnieros} 
\author[ORS]{C.~Nones\fnref{CEA}} 
\fntext[CEA]{Presently at the Service de Physique des Particules, CEA-Saclay, F-91191 Gif sur Yvette, France}
\author[ORS]{E.~Olivieri}
\author[MIB]{G.~Pessina}
\author[INS]{C.~Rusconi}
\author[NVS]{V.N.~Shlegel}
\author[INR]{V.I.~Tretyak}
\author[NVS]{Ya.V.~Vasiliev}

\address[BER]{Lawrence Berkeley National Laboratory, Berkeley, California 94720, USA}
\address[INR]{Institute for Nuclear Research, MSP 03680 Kyiv, Ukraine}
\address[NTS]{Kyiv National Taras Shevchenko University, MSP 03680 Kyiv, Ukraine}
\address[NVS]{Nikolaev Institute of Inorganic Chemistry, 630090 Novosibirsk, Russia}
\address[ORS]{Centre de Spectrom\'etrie Nucléaire et de Spectrom\'etrie de Masse, CNRS and Universit\'e Paris-Sud, F-91405 Orsay, France}
\address[INS]{Universit\`a dell'Insubria,  Dipartimento di Fisica e Matematica, I-22100 Como, Italy}
\address[MIB]{Istituto Nazionale di Fisica Nucleare, Sezione di Milano-Bicocca, I-20126 Milano, Italy}

\begin{abstract}
The search for neutrinoless double $\beta$ decay probes lepton number conservation with high sensitivity and investigates the neutrino nature and mass scale. Experiments presently in preparation will cover the quasi-degeneracy region of the neutrino mass pattern. Probing the so-called inverted hierarchy region requires improved sensitivities and next-generation experiments, based either on large expansions of the present searches or on new ideas. We examine here the potential of a novel technology relying on ZnMoO$_4$ scintillating bolometers, which can provide an experiment with background close to zero in the ton~$\times$~year exposure scale. The promising performance of a pilot detector is presented, both in terms of energy resolution and background control. A preliminary study of the sensitivities of future experiments shows that the inverted hierarchy region is within the reach of the technique here proposed. A realistic phased approach program towards a next-generation search is presented and briefly discussed. 

\end{abstract}


\end{frontmatter}


\section{Introduction and motivations}
\label{sec:intro}

Neutrinoless double $\beta$ decay ($0\nu2\beta$) is a hypothetical rare nuclear transition in which an even-even nucleus changes into an isobar by the simultaneous emission of two electrons and nothing else \cite{DBD-rev}. The observation of this process would imply the violation of the lepton number conservation and definitely new physics beyond the Standard Model, establishing the Majorana nature of neutrinos. If the transition takes place through the so-called mass mechanism, the decay rate is proportional to the square of the effective Majorana mass $m_{\beta \beta}$, that can be determined or at least constrained within the uncertainties of the nuclear matrix elements. This parameter is related to the absolute neutrino mass scale and depends on the three neutrino masses $m_1$, $m_2$ and $m_3$. It is convenient to distinguish three mass patterns: normal hierarchy (NH), where $m_1 < m_2 < m_3$, inverted hierarchy (IH), where $m_3 < m_1 < m_2$, and quasi-degenerate pattern (QD), where the differences between the masses are small with respect to their absolute values. We ignore Nature's choice about the neutrino mass ordering at the moment, but $0\nu2\beta$ has the potential to provide this essential information \cite{DBD-rev,REV-vissani}, given the relationship between $m_{\beta \beta}$ and the three neutrino masses. In fact, if $m_{\beta \beta}$ is measured to be greater than $\approx50$~meV, the QD pattern holds and an allowed range of $m_{min}$ values  can be extracted. On the other hand, if $m_{\beta \beta}$ lies in the range $20-50$~meV, the pattern is likely IH. Eventually, if one determined that $m_{\beta \beta} < 10$~meV but non-vanishing (which is unlikely in a foreseeable future), one would conclude that the NH pattern holds. The process $0\nu2\beta$ is important both for the comprehension of fundamental aspects of neutrino physics and for the solution of hot astroparticle and cosmological issues, intimately related to the neutrino mass scale and nature.

The signature for $0\nu2\beta$ consists of a peak located at the Q-value of the transition in a spectrum of the sum of the energy of the two emitted electrons \cite{DBD-rev}. The current sensitivity to $m_{\beta \beta}$ is around $0.2-0.5$~eV. A much debated claim of evidence in $^{76}$Ge corresponds to $m_{\beta \beta} \approx 0.3$~eV \cite{KLA}. Experiments under commissioning or construction \cite{DBD-rev} can hardly start to explore the IH region, with sensitivities around $0.1-0.05$~eV. In order to deeply analyze it, relevant expansions or improvements of current experiments are needed, or new technologies need to be developed. In this work, we propose to use scintillating bolometers for a frontier $0\nu2\beta$ experiment focused on the study of $^{100}$Mo and capable to explore the IH range.

A bolometric detector \cite{REV-bolo} consists of an energy absorber, in the form of a single crystal, equipped with a temperature sensor. The signal, collected at very low temperatures (typically $< 20$ mK for large bolometers), consists of a temperature rise of the whole detector determined by a nuclear event. Due to the high energy resolutions and to the wide flexibility in the choice of the detector material, this approach is well tailored to the demands of a sensitive $0\nu2\beta$ experiment, based on the so-called {\em source = detector} technique \cite{fio-nii}. Ultra-pure crystals up to $100-1000$~g can be grown with interesting materials, containing appealing candidates. The single crystal module can be multiplied in order to achieve total masses of the order of $100-1000$~kg, necessary to explore the IH region. The CUORE experiment, under commissioning, applies this techonology to the study of the candidate $^{130}$Te in arrays of natural TeO$_2$ bolometers with $\approx 200$~kg isotope mass \cite{CUORE}. 

When the energy absorber in a bolometer scintillates at low temperatures, the simultaneous detection of scintillation light and heat provides a very powerful tool to identify the nature of the interacting particle and therefore to suppress background. In particular, a massive charged particle can be separated from an electron or $\gamma$ due to the different light yield for the same amount of deposited heat, as proposed more than twenty years ago \cite{gonzales-mestres}. In the field of double $\beta$ decay, the first experimental proof of this concept was achieved in 1992 with a CaF$_2$ scintillating bolometer, developed as a pilot device for the search for $0\nu2\beta$ in $^{48}$Ca \cite{scint-milano}. Recently, this approach was proposed to study $0\nu2\beta$ of $^{82}$Se (LUCIFER project) with the help of ZnSe crystals \cite{LUCIFER}. In non-scintillating materials relevant for $0\nu2\beta$, the particle identification can be achieved through the detection of the much weaker Cherenkov light \cite{taba, cere}. In the current technology, the light is detected by thin dedicated bolometers facing the main one. 

Scintillating bolometers containing a double $\beta$ decay emitter with a $Q$-value above the 2615 keV $^{208}$Tl $\gamma$ line are very promising devices for a future $0\nu2\beta$ experiment. In fact, the energy region extending above 2615 keV is almost free from $\gamma$ background due to natural radioactivity, but is dominated by $\alpha$ particles, as the experience brought by Cuoricino and CUORE-R\&D clearly shows \cite{BKG-alpha}. Hence derives the power and the potential of scintillating bolometers. They offer a reasonable freedom in the choice of the candidate, that can be selected for its high $Q$-value (such as the here proposed $^{100}$Mo, for which $Q$=3034.40(17) keV \cite{Q-Mo}), with the additional advantage that $\alpha$ particles can be recognized and rejected. The performance of the protoype here presented (Section \ref{sec:proto}) and a detailed background evaluation in a realistic set-up (Section\ref{sec:design}) show that an experiment with background close to zero in the ton~$\times$~year scale exposure is viable with the proposed technique (Section \ref{sec:prosp}).

\section{A ZnMoO$_4$ scintillating bolometer prototype}
\label{sec:proto}

The validation of the approach here discussed has been obtained through the successful operation of a ZnMoO$_4$ scintillating bolometer prototype. ZnMoO$_4$ is an attractive material for the proposed application, as it is an intrinsic scintillator with a high molybdenum content (43\% in mass).  It forms white tetragonal crystals with a density of 4.3 g/cm$^3$. 
\begin{figure}[t]
\centering
\includegraphics[width=0.7\columnwidth]{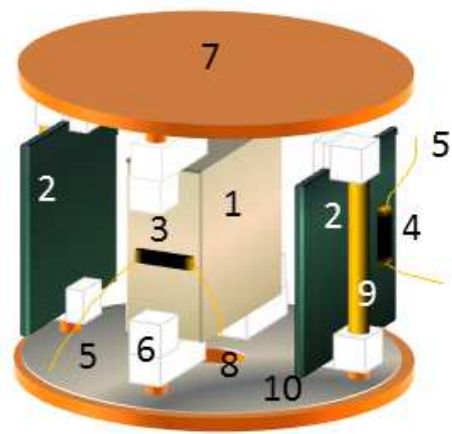} 
\caption{Schematic view of the detector prototype, after removal of its  cylindric Cu container: (1) ZnMoO$_4$ crystal ($\approx$15~$\times$~15~$\times$~5~mm); (2) Square Ge light detectors (15 mm side and 0.5 mm thickness); (3) NTD thermistor glued on the ZnMoO$_4$ crystal; (4) NTD thermistor glued on the Ge slab; (5) ultra-thin gold wires ($\oslash$50~$\mu$m) to read out signals form the thermistors; (6) PTFE holder of the ZnMoO$_4$ crystal; (7) Copper support of the detector (acting as heat sink); (8) $\alpha$ source; (9) PTFE/brass support of the Ge light detector; (10) light reflector.}
\label{fig:detector}
\end{figure}

ZnMoO$_4$ crystals of high quality were developed in the Nikolaev Institute of Inorganic Chemistry (Novosibirsk, Russia). The crystal synthetization was preceded by a deep chemical purification of molybdenum (in the form of MoO$_3$), with the objective to get rid of metal contamination, which may jeopardize the scintillation yield. ZnMoO$_4$ crystals up to 25 mm in diameter and 60 mm in length were grown by the low-thermal-gradient Czochralski technique in a platinum crucible with a size of $\oslash40\times100$ mm \cite{Gala10}. Optical characterization of the samples, described in more details in \cite{tech}, showed a broad emission spectrum under X-ray irradiation, peaking at around 625~nm at 8~K. The optical quality of the crystals resulted largely improved with respect to previous samples.

After these preliminary investigations, a detector prototype was designed, fabricated and cooled down to test its bolometric behaviour and the $\alpha$/$\beta$ rejection factor. The tests have been performed aboveground in the cryogenic laboratory of the University of Insubria (Como, Italy) and in the Centre de Spectrom\'etrie Nucl\'eaire et de Spectrom\'etrie de Masse (Orsay, France). The realized detector (see Figure~\ref{fig:detector}) consists of a rectangular ZnMoO$_4$ crystal with a mass of 5.07 g, faced by two light-detecting ultrapure Ge thin slabs. The three bolometers are surrounded by a highly reflective polymeric multilayer foil (Radiant Mirror Film VM2000/VM2002 from 3M). The thermal signals from the ZnMoO$_4$ crystal and the two Ge slabs were read out by three nominally identical sensors, consisting of neutron transmutation doped (NTD) Ge thermistors \cite{NTD}, with a mass of $\approx10$~mg. The thermistor resistances and sensitivities are tuned for an optimal operation in the $20-30$~mK range.

In some runs, radioactive sources were placed in the vicinity of the detector for calibration purposes. In particular, a collimated $^{241}$Am source, characterized by a main $\alpha$ line at $5.48$~MeV and an intense $\gamma$ line at 59~keV, illuminated the ZnMoO$_4$ crystal at the center of a 15~$\times$~5~mm face. Two weak $^{55}$Fe sources, providing X-rays at 5.9~keV and 6.4~keV, irradiated the external side of each Ge slab.

The detector was operated at various base temperatures -- between 25 and 32 mK -- in two dilution refrigerators. The typical NTD thermistor resistances at the operation points which provided the best performance were of the order of 1~M$\Omega$, with bias currents ranging between 2 and 5~nA. For the electronic readout, the NTD Ge thermistors were connected to low noise voltage amplifiers \cite{Pess}, which presented a cold stage inside the cryostat in Orsay only. The full waveforms of the signal were acquired and registered, and the method of the optimum filter \cite{OF} was applied in order to maximize the detector energy resolution in an off-line analysis. 

\begin{figure}[t]
\centering
\includegraphics[width=\columnwidth]{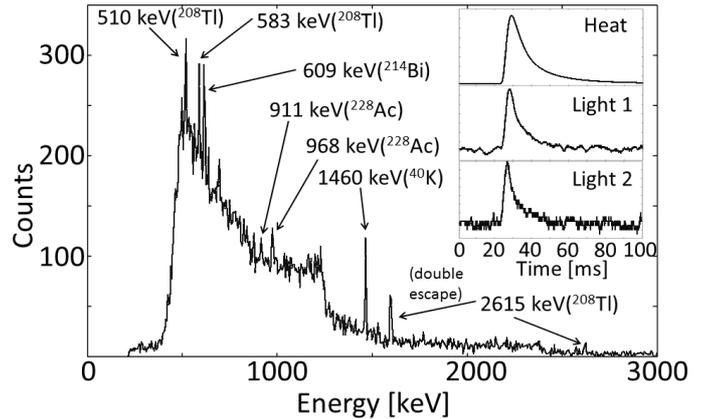} 
\caption{Background $\gamma$ spectrum collected with the heat channel of the ZnMoO$_4$ bolometer, showing several environmental $\gamma$ peaks. In the inset, simultaenous heat and light pulses for the same event are shown, with an arbitrary amplitude scale.}
\label{fig:gamma}
\end{figure}

The signals from the ZnMoO$_4$ crystal (the heat channel) corresponded to voltage pulses with an amplitude of $\approx 200$~$\mu$V for 1 MeV deposited energy at 25~mK. The signal time structure was characterized by $\approx 1$~ms risetime and $\approx 10$~ms decaytime (respectively from 10\% to 90\% and from 90\% to 30\% of the signal maximum amplitude). This time behaviour is consistent with the thermal network representing the detector, and corresponds to what commonly observed in dielectric macrobolometers operated in similar conditions. The overall performance of the detector is compatible with the operation of large crystals (of the order of hundreds of grams) in the temperature range $10-15$~mK, as in the case of TeO$_2$ bolometers in the CUORE experiment. The amplitude of the signals from the two Ge light detectors was $\approx 1 - 2 $~$\mu$V for 1 keV deposited energy. The pulses were slightly faster with respect to the heat channel ($\approx 0.9$~ms risetime and $\approx 7$~ms decaytime). The presence of two light detectors was very useful to identify event classes through triple coincidences (see inset in Figure~\ref{fig:gamma}).

The $\gamma$ calibrations of the ZnMoO$_4$ detector (heat channel) were performed using the low energy $\gamma$ line provided by the internal $^{241}$Am source and external $\gamma$s emitted by a $^{226}$Ra source or due to environmental radioactivity. In the Como set-up, where the cryostat is unshielded, a background measurement showed many typical environmental $\gamma$ lines (Figure~\ref{fig:gamma}). The intrinsic energy resolution of the heat channel (corresponding to the FWHM baseline fluctuations) is of the order of 800 eV. The energy resolutions on the lines are worse (this is commonly observed in macro-bolometers), but they are always excellent, of the order of 1.3 keV FWHM at the 59 keV $^{241}$Am line and, at higher energies, of 3.8 keV FWHM at the 2615 keV environmental $^{208}$Tl line. This performance shows that ZnMoO$_4$ is a very promising material for high energy resolution bolometers. The position of the main $\approx 5.5$~MeV $\alpha$ peak of the $^{241}$Am source on a $\gamma$-calibrated energy scale shows that the phonon yield of an $\alpha$ particle is larger than that of a $\gamma$ or a $\beta$ by a factor $\approx 1.105$. A phonon pulse formation model able to explain this result is lacking for the moment. However, this behaviour was observed in ZnSe bolometers as well \cite{ZnSe-MI}.  

\begin{figure}[htp]
\centering
\includegraphics[width=\columnwidth]{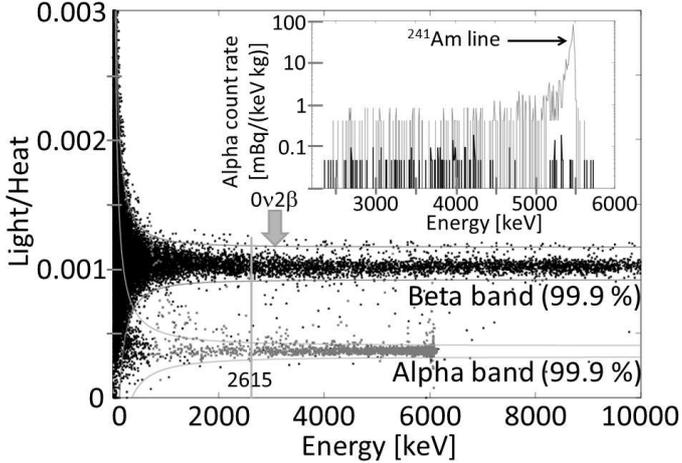} 
\caption{The ratio of the light-to-heat energy for the same event (using the light detector with better performance) is plotted against the $\gamma$-calibrated heat energy. The $\alpha$ band (populated by a $^{241}$Am source) is neatly separated from the $\beta$ band (containing also $\gamma$s and muons). Lines define 99.9\% c.l. regions. Grey points indicate $\alpha$ events selected with pulse shape discrimination in the heat channel. In an underground set-up, the $\beta$ band would not be populated above 2615 keV, due to a substantial reduction of the cosmic ray flux. In the inset, $\alpha$-calibrated energy spectra containing only $\alpha$ events are reported, with -- grey lines -- and without -- black lines -- the $^{241}$Am source.}
\label{fig:Q-plot}
\end{figure}

Observing the time coincidences between signals collected with the Ge slabs and the ZnMoO$_4$ crystal (see the inset in Figure~\ref{fig:gamma}), and thanks to the energy calibration of the heat and light channels, it was possible to deduce the light yield for $\alpha$ and $\beta$ particles. It resulted respectively of 2.08 keV/MeV for $\beta$s, $\gamma$s and cosmic muons, and of 0.42 keV/MeV for $\alpha$ particles. These values, consistent with previous measurements \cite{GiroZMO}, take into account the light collected by both light detectors (which is the same in both of them for the same event in the ZnMoO$_4$ crystal within the light detector energy resolution), but do not account for the light collection efficiency. Therefore, they are actually inferior limits. Even considering this, it is clear that ZnMoO$_4$ is a very poor intrinsic scintillator. However, the light output is largely sufficient to perform efficiently the $\alpha$/$\beta$ discrimination. An $\alpha$ rejection factor much better than 99.9 \% has been determined at the energy of interest for $^{100}$Mo $0\nu2\beta$. The discrimination capability of the detector is appreciable in Figure~\ref{fig:Q-plot}.

As already observed in previous works \cite{ZnSe-MI, GiroZMO, GiroCD, novel}, in scintillating bolometers (and in particular in molybdates and tungstates) $\alpha$ and $\beta$ particles can be distinguished using the heat channel only through pulse shape discrimination. This is the case for our prototype too. The $\alpha$ particle signals were faster than $\gamma$/$\beta$/muon signals by a few \% both in the rise and in the decay times. Once again, we are not able to explain this feature, a model lacking for the phonon pulse formation. However, it can be efficiently used for rejection purposes. We have constructed the sum over the sampling points of the squares of the differences between each pulse and the average pulse (once normalized). The distribution of this parameter allows to achieve the same information gathered by the simultaneous detection of heat and light (see Figure~\ref{fig:Q-plot}). 

In order to determine the intrinsic contamination of the crystal, we have performed a run without the $^{241}$Am $\alpha$ source. The $\alpha$ background spectrum is reported in the inset of Figure~\ref{fig:Q-plot} and compared with the $\alpha$ calibration spectrum. Over an exposure of 25.6~g~$\times$~days, we observed hints for a surface contamination only, probably in $^{210}$Po or $^{210}$Pb, and detectable through a tail below 5.5~MeV. This phenomenon is observed in other bolometric experiments like Cuoricino \cite{Cuoricino} and EDELWEISS \cite{EDW}. Since there is no evidence of internal $\alpha$ peaks, it is possible to set limits on the contamination of several nuclides belonging to the natural radioactivity chains looking at their first $\alpha$ decay. When the expected $\alpha$ peak is above 5.5 MeV, we have used both background and calibration spectrum in order to increase the exposure up to 43.1~g~$\times$~days. For the isotopes $^{232}$Th, $^{228}$Th, $^{227}$Ac, $^{238}$U, $^{234}$U, $^{230}$Th, $^{226}$Ra and $^{212}$Bi an upper limit (at 90\% c.l.) of respectively 3.3 mBq/kg, 1.1 mBq/kg, 1.1 mBq/kg, 2.7 mBq/kg, 1.1 mBq/kg, 1.7 mBq/kg, 1.1 mBq/kg and 0.64 mBq/kg can be set. A substantial improvement was observed with respect to a ZnMoO$_4$ crystal tested in a previous work \cite{GiroZMO}, in particular for the isotope $^{226}$Ra, which exhibited an internal activity -- now absent --  of $8.1 \pm 0.3$~mBq/kg. Above 5.5 MeV and therefore out of the region affected by the common contaminants $^{210}$Po and/or $^{210}$Pb, the $\alpha$ background spectrum is empty. In particular, the result on  $^{212}$Bi (expected peak at 6208 keV) allows to set a limit on the internal activity of its dangerous daughter $^{208}$Tl (see next Section) at 0.23 mBq/kg. This shows that the molybdenum purification method and the whole crystallization procedure are quite promising for a future $0\nu2\beta$ experiment. 
 
\section{Background discussion of future ZnMoO$_4$ experiments}
\label{sec:design}

In the region where the $^{100}$Mo $0\nu2\beta$ signal is expected, the energy-degraded $\alpha$ continuum is presently the dominant background sources in bolometric experiments, at the level of $0.1-0.05$~counts/(keV kg y) \cite{BKG-alpha}. An $\alpha$ rejection factor better than 99.9\% would allow to reduce this component to less than 10$^{-4}$~counts/(keV kg y). This implies virtually zero background in the ton~$\times$~year scale exposure. However, when designing a next-generation experiment, a more careful investigation is mandatory. A realistic Monte Carlo simulation of the background in the region of interest was performed. The ingredients for this simulation are the structure of the single detector module and of the detector array, and the type and intensity of the background sources. The experimental results presented in the previous Section provide a guideline for the design of a future set-up containing radiopure ZnMoO$_4$ crystals. 

The single module we propose here is based on a ZnMoO$_4$ crystal with the shape of a hexagonal prism 4 cm high and with the maximum diagonal 6 cm long. This corresponds to a detector mass of $\approx 400$~g. We propose a hexagonal shape as there is experimental evidence \cite{hexag}, supported by simulations, that this configuration (with diffused rather than optical grade surfaces) improves the light collection uniformity and the efficiency by a factor between 20\% and 30\% with respect to a cylinder with a diameter equal to the hexagonal diagonal. Other possibilities can be considered, like the intermediate case of an octagonal prism. Each ZnMoO$_4$ crystal will be held by six PTFE blocks similar to those used in the prototype discussed here, with a mass of about 0.5~g each. It will be surrounded by a cylindrical Cu holder with the internal surface covered by a reflective foil. One extremity of the Cu holder will carry a light detector consisting of a Ge disk with 6~cm diameter and 0.5~mm thickness, facing one base of the prism. This is the baseline option, on which the Monte Carlo is based. Another option foresees two identical light detectors at the two extremities, in analogy with the configuration of the prototype here presented. We assume that the single modules are arranged in towers. In order to include the benefits of the array granularity in the Monte Carlo, which enables coincidence cuts, we have designed a test structure with 7 tightly packed towers  (in a top view, 6 towers sit at the vertices of a hexagon and the 7$^{th}$ at its center). Each tower contains 7 scintillating bolometers, providing a global approximate cylindrical array with 49 modules. Other similar structures are conceivable with comparable results. 

In the Monte Carlo simulation, we have considered the following possible sources:
\begin{enumerate}
\item Internal $^{208}$Tl and $^{214}$Bi in ZnMoO$_4$;
\item Contamination of the crystal support and surrounding materials by $^{228}$Th and $^{226}$Ra;
\item Cosmogenically produced radionuclides inside the ZnMoO$_4$ crystal scintillators and the Cu details of the set-up.
\end{enumerate}
The choice of the first two sources is dictated by the following considerations. Although in the $^{238}$U and $^{232}$Th chains there is no intense $\gamma$ line above 2615~keV, there are however two $\beta$ active nuclides -- $^{214}$Bi and $^{208}$Tl -- with $Q$-values greater than 3 MeV (respectively, 3.270 and 4.999 MeV). Thus, these two isotopes produce $\gamma$s and $\beta$s that are energetic enough to give events around 3 MeV on the whole if they are partially absorbed together. A particular dangerous case takes place when the contamination is either in the ZnMoO$_4$ crystal itself or in the surface of the material facing it; in fact, this allows $\beta$ electrons to deposit efficiently energy in the crystal. External $\gamma$s from $^{208}$Tl are dangerous too, because the 2615 keV photon is always in close cascade with others which have enough energy (511 keV, 583 keV and 860 keV) to populate the 3 MeV region when absorbed in coincidence with it. Anticoincidences help in reducing this component. We have to remark also the presence of weak $^{214}$Bi lines (B.R.$\approx 0.15$\% in the $^{238}$U chain) above 2615~keV. 

The first source in the above list addresses directly the internal contamination of $^{214}$Bi and $^{208}$Tl. The second source assumes that these isotopes are generated externally by their predecessors (supposed in secular equilibrium with the daughters) $^{228}$Th and $^{226}$Ra, whose typical concentration limits in ultra-radio-pure Cu and PTFE are well known. The ultrapure Ge light detector is not a dangerous source of radioactivity, because any conceivable process originating in it that could deposit around 3 MeV in the ZnMoO$_4$ crystal would give also an abnormally high signal in the light detector itself. 

The background generated by the aforementioned sources was simulated with the help of the GEANT4 \cite{GEANT4} package and the event generator DECAY0 \cite{DECAY0}, assuming conservatively an energy resolution of 6 keV at 3 MeV and a threshold at 20~keV, compatible with the tests reported here and elsewhere \cite{GiroZMO}.
\begin{table}[h]
\caption{Monte Carlo simulated background contributions in a future search for $0\nu2\beta$ of $^{100}$Mo based on scintillating bolometers of ZnMoO$_4$ (ZMO). BKG denotes the specific background rate at the $Q$-value of $^{100}$Mo $0\nu2\beta$ decay.}
\centering
\begin{tabular}{lcc}
\hline
Source of  & Activity & BKG  \\
background & [$\mu$Bq/kg] & [counts/(keV kg y)]\\
\hline
$^{208}$Tl in ZMO   & 10    & $3.2\times10^{-3}$ \\
$^{214}$Bi in ZMO  & 10    & $3 \times10^{-8}$ \\
$^{228}$Th in Cu  & 20    & $1.6\times10^{-5}$ \\
$^{226}$Ra in Cu  & 70    & $1.3\times10^{-7}$ \\
$^{228}$Th in PTFE    & 100     & $2\times10^{-7}$ \\
$^{226}$Ra in PTFE    & 60    & $< 10^{-9}$ \\
$^{56}$Co in ZMO  & 0.06 & $1.8\times10^{-5}$ \\
$^{56}$Co in Cu  & 0.02 & $8\times10^{-6}$ \\
$^{88}$Y  in ZMO & 0.3  & $7\times10^{-7}$ \\
\hline
Total                    &      & $3.2\times10^{-3}$ \\
\hline
\end{tabular}
\label{tab:simul}
\end{table}

The results of the simulation are summarized in Table~\ref{tab:simul}, which reports also the assumed contamination levels. A rejection factor of 99.9\% was applied every time that a $\beta$ event is emitted in sequence with an $\alpha$ event within the time resolution of the bolometer, as for the $^{212}$Bi-$^{212}$Po and $^{214}$Bi-$^{214}$Po cascades. For copper and PTFE, the radioactive impurity concentration was assigned using measurements and limits reported in literature \cite{cont-Cu, cont-PTFE}. For ZnMoO$_4$ crystals, contamination by $^{228}$Th and $^{226}$Ra were assumed to be at the level of 0.01 mBq/kg, which is realistic looking at the experience of other bolometric set-ups and considering the promising purification methods established for $^{100}$Mo and here applied to a test bolometer. A comparable level of radio-purity was already reached for 
ZnWO$_4$ crystal scintillators \cite{cont-ZWO}, produced with the same initial zinc oxide as that used for the synthetization of the ZnMoO$_4$ crystals. Possible cosmogenically induced activity in ZnMoO$_4$ was estimated with the COSMO code \cite{COSMO} assuming 3 months of activation on the Earth surface and 1 yr cooling underground. We have found two dangerous nuclides with $Q_{\beta}>3035$ keV: $^{56}$Co ($Q_{\beta}=4566$~keV, $T_{1/2}=77.27$~days) and $^{88}$Y ($Q_{\beta}=3623$~keV, $T_{1/2}=106.65$ days). $^{88}$Y can also appear as daughter of cosmogenic $^{88}$Zr ($T_{1/2}=83.40$ days). 

The only really dangerous component in Table~\ref{tab:simul} is the internal disintegration of $^{208}$Tl, which contributes at the level of $\approx3 \times 10 ^{-3}$ counts/(keV kg y). However, $^{208}$Tl has a short half-life (3.053~min), and can only be produced internally by the $\alpha$ decay (B.R. 35.9~\%) of $^{212}$Bi, belonging to the $^{232}$Th chain. Given the high energy resolution of the detector and its ability to identify $\alpha$s, every time that an $\alpha$ count is compatible with a $^{212}$Bi $\alpha$ decay (i.e. contained in a sharp peak at 6207 keV), all the acquired pulses can be discarded for a time gate of $\approx 5$~times the half life of $^{208}$Tl, i.e. about 15 minutes. This will reduce the corresponding background by a factor $\exp(-15 \cdot \ln(2)/3.053)\approx1/30$, with a realistic abatement down to $\approx1 \times 10 ^{-4}$ counts/(keV kg y). The live time reduction due to this operation is negligible (of the order of 1\%), since a $^{212}$Bi $\alpha$ decay is expected every $\approx1500$ minutes in a 400~g crystal with the assumed contamination.

The background caused by neutrons and muons has to be considered as well. For these sources, we refer to the estimation made by the CUORE collaboration for a high granularity bolometric set-up in an underground cryostat, in a configuration similar to the one proposed here \cite{CUORE-neu}. The neutron contribution, using near detector anticoincidence cuts, is foreseen to be well below $10^{-4}$ counts/(keV kg y). We point out that cross-sections for thermal neutron interactions on Zn, $^{100}$Mo and O isotopes do not exceed 6.8 barn ($^{67}$Zn). The muon contribution, with coincidence cuts, is of the order of $10^{-4}$ counts/(keV kg y), but it can be substantially reduced with a muon veto system.

A further background source to be considered is represented by the standard two-neutrino double $\beta$ decay ($2\nu2\beta$) of $^{100}$Mo ($T_{1/2}=7.11\times10^{18}$~y \cite{2nu}). This process produces a continuum in the two-electron sum energy spectrum extending from zero up to the $Q$-value, and peaked roughly at $1/3$ of the $Q$-value itself \cite{DBD-rev}. Due to the excellent energy resolution of the devices here proposed, the high-energy $2\nu2\beta$ spectrum tail is not harmful. However, the random coincidences of two $2\nu2\beta$ events can produce appreciable background. In a 400~g crystal enriched in $^{100}$Mo at 97\%, the total $2\nu2\beta$ rate is 3.15~mBq. Considering the shape of the pile-up spectrum, it is possible to compute the background due to this effect, which scales as the square of the total rate. It is $\approx2.7\times10^{-4}\cdot$[$T_R$/1 ms]~counts/(keV kg y), where $T_R$ is the pulse-pair resolving time. In bolometers of hundreds of grams, $T_R$ would be of the order of tens of milliseconds, since this is the typical heat signal risetime \cite{Cuoricino}. However, the fast risetime of the light detector here observed would be preserved in large bolometers since the Ge energy absorber would have a mass of only a few grams. This allows to take safely $T_R$ of the order of 1~ms, and therefore the $2\nu2\beta$ background contribution around $3\times10^{-4}$ counts/(keV kg y). This level would be 100 times less, and therefore negligible, in a crystal containing natural molybdenum. 
  
We can therefore assume a total background rate in the region of interest of $\approx4 \times 10^{-4}$ counts/(keV kg y) for enriched crystals, corresponding to 2.4 counts/(y ton) in a 6 keV window centered at the $0\nu2\beta$ peak position. A factor 4 less can be assumed for natural crystals.
\begin{table}[t]
\caption{Sensitivity (at 90\% c.l., Bayesian approach) of experiments based on ZnMoO$_4$ scintillating bolometers searching for $0\nu2\beta$ of $^{100}$Mo. The assumed background is discussed in the text. The live time is 5 y and the energy window 6~keV, compatible with the performed tests. The enrichment level is 97\% (except for the third option, where natural molybdenum is considered) and the efficiency is 90\%, compatible with the single module structure. The range in $m_{\beta \beta}$ takes into account three different approaches to the evaluation of the nuclear matrix elements (QRPA \cite{QRPA-1, QRPA-2}, ISM \cite{ISM}, IBM-2 \cite{IBM}).}
\centering
\begin{tabular}{ccccc}
\hline
\ & Number of& Total  & Half-life  & $m_{\beta \beta}$ \\
Option & $\approx400$~g & isotope  & sensitivity &  sensitivity  \\
\ & crystals & mass [kg] & [10$^{25}$~y] & [meV] \\
\hline
$(1)$ & 4 & 0.676 & 0.53 & $167-476$ \\
$(2)$ & 40 & 6.76 & 4.95 & $55-156$ \\
$(3)$ & 2000 (nat.) & 33.1 & 15.3 & $31-89$ \\
$(4)$ & 2000 & 338 & 92.5 & $13-36$ \\
\hline
\end{tabular}
\label{tab:exp}
\end{table}

\section{Prospects and conclusions}
\label{sec:prosp}
The study of a ZnMoO$_4$ scintillating bolometer prototype and the simulation of the background for a close-packed array of such devices allow us to evaluate the sensitivities of future $0\nu2\beta$ searches in realistic configurations, as reported in Table~\ref{tab:exp}. We consider both the use of enriched and natural material, given the reasonably high isotopic abundance of $^{100}$Mo (9.7\%). For immediate next tests, about 1~kg of $^{100}$Mo belonging to the Kyiv Institute for Nuclear Research is available and an enriched 200~g ZnMoO$_4$ crystal is in preparation. The low-thermal-gradient Czochralski technique minimizes the loss of material during crystal growth. In case of a positive outcome of the 200~g test, the available isotope could be embedded in four large crystals of about $400-500$~g each and measured in one of the existing underground low-activity dilution fridges in Gran Sasso or Modane underground laboratories (option (1) in Table~\ref{tab:exp}). This pilot experiment would have sensitivities comparable to present searches and would constitute the general testbench for scaling up this technology. A further step could exploit existing enriched material, of the order of several kg. This search (option (2) in Table~\ref{tab:exp}) would approach the IH region and could be housed by the former Cuoricino fridge. A major extension is required to substantially cover the IH mass pattern. Molybdenum can be enriched by centrifugation at reasonable prices (of the order of $50-100$~\$/g) and with reasonable throughput. The production of $\approx350$~kg of $^{100}$Mo fits the time and budget scale of a large next-generation $0\nu2\beta$ search. Bolometric masses of the order of 1~ton could be housed, after the completion of the TeO$_2$ program, by the CUORE dilution refrigerator (under construction) or by the EURECA dilution refrigerator (under design), which could share dark matter and $0\nu2\beta$ searches (options (3) and (4) in Table~\ref{tab:exp}). In addition to the investigation of $^{100}$Mo, a large experiment could detect also the standard two neutrino double electron capture in $^{64}$Zn. This isotope has high isotopic abundance ($\approx49\%$). A value of $T_{1/2}=(1.2\pm0.2)\times10^{25}$~y was predicted for this process \cite{Zn64}, but it should be noted that other calculations give $T_{1/2}$ values one order of magnitude higher. Finally, a detector with a large mass of $^{100}$Mo can perform real-time spectrometry of low-energy neutrinos from different sources (sun, supernovae, etc.) thanks to the low energy threshold (168 keV) and large cross section for inverse $\beta$ decay \cite{LEnu}.

In conclusion, the work reported in the present paper shows that arrays of ZnMoO$_4$ scintillating bolometers are viable candidates for a next-generation $0\nu2\beta$ experiment, capable to explore the IH region. A program for a phased approach to a large scale search is feasible with existing means and will be pursued in the immediate future.

\section{Aknowledgements}

The work of F.A.~Danevich was supported by a Cariplo Foundation fellowship organized by the Landau Network - Centro Volta (Como, Italy). The group from the Institute for Nuclear Research (Kyiv, Ukraine) was supported in part through the Project ``Kosmomikrofizyka-2'' (Astroparticle Physics) of the National Academy of Sciences of Ukraine. The light detectors have been realized within the project LUCIFER, funded by the European Research Council under the EU Seventh Framework Programme (ERC grant agreement n. 247115).






\begin{thebibliography}{00}


\bibitem{DBD-rev} F.T.~Avignone III, S.R. Elliott, J. Engel, Rev. Mod. Phys. 80 (2008) 481; \\ A. Giuliani, Acta Phys. Polon. B 41 (2010) 1447. 
\bibitem{REV-vissani} A. Strumia, F. Vissani, arXiv:hep-ph/0606054v3.
\bibitem{KLA} H.V. Klapdor-Kleingrothaus, et al., Phys. Lett. B 586 (2004) 198; \\ H.V. Klapdor-Kleingrothaus, I.V. Krivosheina, Mod. Phys. Lett. A 21 (2006) 1547.
\bibitem{REV-bolo} A. Giuliani, Physica B 280 (2000) 501.
\bibitem{fio-nii} E. Fiorini, T.O. Niinikoski, Nucl. Instr. Meth. 224 (1984) 83.
\bibitem{CUORE} C. Arnaboldi, et al., Astropart. Phys. 20 (2003) 91; \\
C. Arnaboldi, et al., Nucl. Instr. Meth. A 518 (2004) 775.
\bibitem{gonzales-mestres} L. Gonzalez-Mestres, D. Perret-Gallix, in Low Temperature Detectors for Neutrinos and Dark Matter - II, Proceedings of LTD-2, Annecy, France (1988), Editions Fronti\`eres.
\bibitem{scint-milano} A. Giuliani, S. Sanguinetti, Mater. Sci. Eng. R11 (1993) 1; \\ A. Alessandrello, et al., Phys. Lett. B 420 (1998) 109. 
\bibitem{LUCIFER} A. Giuliani, et al., ``LUCIFER: an experimental breakthrough in the search for neutrinoless double beta decay'', Proceedings of the 5th International BEYOND 2010 Conference, Cape Town, South Africa (2010), World Scientific.
\bibitem{taba} T. Tabarelli de Fatis, Eur. Phys. J. C 65 (2010) 359. 
\bibitem{cere} J.W. Beeman, et al., arXiv:1106.6286v1.
\bibitem{BKG-alpha} M. Pavan, et al., Eur. Phys. J. A 36 (2008) 159. 
\bibitem{Q-Mo} S.~Rahaman, et al., Phys. Lett. B 662 (2008) 111.
\bibitem{Gala10} E.N.~Galashov, et al., Functional Materials 17 (2010) 504.
\bibitem{tech} F.A.~Danevich, et al., ``Improved ZnMoO$_4$ crystal 
scintillators for cryogenic double $\beta$ decay experiments'', in preparation for Nucl. Instr. Meth. A.
\bibitem{NTD} E. Haller, J. Appl. Phys. 77 (1995) 2857.
\bibitem{Pess} C. Arnaboldi, et al., IEEE Trans. Nucl. Sci. 49 (2002) 2440.
\bibitem{OF} E. Gatti, P.F. Manfredi, Rivista Nuovo Cimento 9 (1986) 1.
\bibitem{ZnSe-MI} C. Arnaboldi, et al., Astropart. Phys. 34 (2011) 344.
\bibitem{GiroZMO} L.~Gironi, et al., JINST 5 (2010) 11007.
\bibitem{GiroCD} C. Arnaboldi, et al., Astropart. Phys. 34 (2010) 143. 
\bibitem{novel} C. Arnaboldi, et al., Astropart. Phys. 34 (2011) 797.
\bibitem{Cuoricino} E. Andreotti, et al., Astropart. Phys. 34 (2011) 822.
\bibitem{EDW} S. Fiorucci, et al., Astropart. Phys. 28 (2007) 143. 
\bibitem{hexag} F.A.~Danevich, et al., ``Optimization of light collection from ZnWO$_4$ crystal scintillators in cryogenic experiments'', in preparation for Nucl. Instr. Meth. A.
\bibitem{GEANT4} S.~Agostinelli, et al., Nucl. Instr. Meth. A 506 (2003) 250; \\
J.~Allison, et al., IEEE Trans. Nucl. Sci. 53 (2006) 270.
\bibitem{DECAY0} O.A.~Ponkratenko, et al., Phys. At. Nucl. 63 (2000) 1282.
\bibitem{cont-Cu} C.~Dorr, H.V.~Klapdor-Kleingrothaus, Nucl. Instr. Meth. A 513 (2003) 596.
\bibitem{cont-PTFE} E.~Aprile, et al., Astropart. Phys. 35 (2011) 43.
\bibitem{cont-ZWO} P.~Belli, et al., Nucl. Instr. Meth. A 626 (2011) 31.
\bibitem{COSMO} C.J.~Martoff, P.D. Lewin, Comp. Phys. Comm. 72 (1992) 96.
\bibitem{CUORE-neu} F. Bellini, et al., Astropart. Phys. 33 (2010) 169.
\bibitem{2nu} R. Arnold, et al., Phys. Rev. Lett. 95 (2005) 182302 .
\bibitem{QRPA-1} F. Simkovic, et al., Phys. Rev. C 79 (2009) 055501.
\bibitem{QRPA-2} O. Civitarese, J. Suhonen, J. Phys.: Conf. Ser. 173 (2009) 012012.
\bibitem{ISM} J. Menendez, et al., J. Phys.: Conf. Ser. 267 (2011) 012058.
\bibitem{IBM} J. Barea, F. Iachello, Phys. Rev. C 79 (2009) 044301.
\bibitem{Zn64} E.W.~Grewe, et al., Phys. Rev. C 77 (2008) 064303.
\bibitem{LEnu} K.G.~Balasi, et al., J. Phys. Conf. Ser. 203 (2010) 012101.

\end{thebibliography}



\end{document}